\documentclass[12pt, draftclsnofoot, onecolumn]{IEEEtran}
\usepackage{graphicx}
\usepackage{color}
\usepackage{graphicx}  
\usepackage{epsf}
\usepackage{amsmath, amsfonts}
\usepackage{latexsym}
\usepackage{subfigure}
\usepackage{multirow}
\usepackage{multicol}
\usepackage{amssymb}
\usepackage{cite}
\usepackage{float}
\usepackage{algorithm}
\usepackage{algpseudocode}
\usepackage{lipsum}

\newtheorem{lemma}{\textbf{Lemma}}
\newtheorem{corollary}{\textbf{Corollary}}
\newtheorem{theorem}{Theorem}
\newtheorem{proposition}{Proposition}
\newtheorem{remark}{Remark}
\newtheorem{definition}{Definition}
\newcommand\numberthis{\addtocounter{equation}{1}\tag{\theequation}}
\newcommand{\argmax}{\operatornamewithlimits{argmax}}

\allowdisplaybreaks

\begin{document}

\title{Optimal Caching Placement for Wireless Femto-caching Network}
\author{Jaeyoung~Song,~\IEEEmembership{Student~Member,~IEEE,}
        Hojin~Song,~\IEEEmembership{Student~Member,~IEEE,}
        and~Wan~Choi,~\IEEEmembership{Senior~Member,~IEEE}
\thanks{This paper was presented in part at the IEEE International Conference on Communication, London, 2015 \cite{icc}}.\thanks{J. Song, H. Song, and W. Choi are with School of Electrical Engineering,
        Korea Advanced Institute of Science and Technology (KAIST), Daejeon
        305-701, Republic of Korea (e-mail: wchoi@kaist.edu).}}

\maketitle

\vspace{-0.2in}
\begin{abstract}
 This paper investigates optimal caching placement for wireless femto-caching network. The average bit error rate (BER) is formulated as a function of caching placement under wireless fading. To minimize the average BER, we propose a greedy algorithm finding optimal caching placement with low computational complexity. Exploiting the property of the optimal caching placement which we derive,  the proposed algorithm can be performed over considerably reduced search space. Contrary to the optimal caching placement without consideration of wireless fading aspects, we reveal that optimal caching placement can be reached by balancing a tradeoff between two different gains: file diversity gain and channel diversity gain. Moreover, we also identify the conditions that the optimal placement can be found without running the proposed greedy algorithm and derive the corresponding optimal caching placement in closed form. 
 \end{abstract}


\IEEEpeerreviewmaketitle

\section{Introduction}\label{Introduction}
The recent spread
of wireless devices has brought heavy data traffic which 
video-streaming requests, such as YouTube, occupy a dominant portion of. Unfortunately, however, current wireless systems have limitations of resources to accommodate this tremendous video
traffic. Discovering new but inexpensive resources is considered a solution to cope with the explosive traffic. In the same vein, memory for data caching  arises as a new resource to exploit in wireless communications \cite{1}. The characteristic of video traffic facilitates utilization of memory to handle huge video traffic; 
a few popular
videos account for the majority of video traffic \cite{2} and hence network traffic to carry the videos to the end users can be 
significantly reduced by storing the top-ranked video files near the users who are likely to request them. 

Although caching has been intensively studied in wired networks, also known as
content delivery network (CDN), it is transparent to wireless segments. Without consideration of wireless aspects, CDN is not enough to provide sufficient data rate to wireless end users. In this context, there have been recent studies figuring out wireless caching. For a given caching placement, several transmission schemes to exploit the given memory contents have been studied in \cite{8,9,11}. Based on appropriate user grouping, when each user requests multiple files, a groupcasting scheme was proposed in \cite{RA4}. With practical Zipf popularity distribution, \cite{RA5} characterized the regimes with different coded multicasting gains. Prefix caching for wireless video streaming  was proposed and optimized in \cite{JP}. Caching was exploited in cooperative multi-point MIMO transmission (CoMP) to  reduce backhaul cost in \cite{CoMP_1,CoMP_2,CoMP_3}. Because where and what to cache determines the performance of wireless caching, caching placement has also attracted research interests, namely femto-caching and device-to-device (D2D) caching. In \cite{RA2},  proactive caching for femto-caching and D2D caching was studied. A caching scheme based on prediction of future demands was proposed in \cite{RA3}. Optimal caching placement is able to minimize the access distance and inter-cell interference in wireless cache networks. In \cite{3, 4},  an optimization problem in terms of the average delay, a function of caching placement, 
was formulated and proved to be NP-complete when each user can access to a different set of femto base stations. With random caching placement for D2D communication, optimal cluster size and the parameter characterizing the random caching distribution were studied in 
\cite{5,6}, where  a square cell was composed of square clusters and D2D communication
was activated if the file requested by user in a cluster was cached in
any device inside the cluster. For D2D coded caching placement, optimal portion of files to store and density of nodes caching the requested file were investigated in \cite{sg_3, 7}, respectively. Combining coded multicasting with D2D communication, \cite{RA1} proposed  caching and delivery schemes which can exploit spatial reuse gain as well as coded multicasting gain. Caching placement optimizing a tradeoff between throughput and outage was explored in \cite{13,14}. When Maximum-distance separable coding is used, optimal storage allocation under a total memory constraint was studied in \cite{Bi,RC1}. It was shown in \cite{RC1} that symmetric allocation achieved asymptotically optimal performance. Replacing outage with delay, a similar tradeoff was revealed in a multi-hop network \cite{15}. 

Although the aforementioned studies on caching placement aim at wireless caching, the wireless channel models in those studies partially address effects of wireless channel fading on caching placement. In particular, constant and identical channel links are assumed in most of the previous works \cite{3,4,5,6,8,9,11,sg_3,7,13,14,15,RA1,RA2,RA3,RA4,RA5}. Although the averaging effect of long file transmission justifies the assumption from a viewpoint of each link, random fading channels might reveal different aspects. For example, 
caching different files according to their popularity (i.e., \emph{file diversity}), which increases the chance to access nearby caching nodes, is reported to be optimal \cite{5} because caching the same file in multiple helpers is redundant under constant and identical channels. However, caching the same file in different helpers might be able to offer \emph{channel diversity} \cite{chd} and hence caching the same file can be rather beneficial in random fading channels. Although channel fading is considered 
in \cite{Bi,CoMP_1,CoMP_2,CoMP_3}, the fundamental tradeoff was not clearly shown since they focused on proposing suboptimal or asymptotic solutions. 

In this context, under random channel fading, this paper revisits and studies the optimal caching placement to minimize average bit error rate (BER) when there exist multiple helpers storing files. We identify the tradeoff between channel diversity and file diversity and derive optimal caching placement optimizing the tradeoff. Our main contributions are as follows:
\begin{itemize}
\item  We propose a greedy algorithm to find the optimal caching placement in terms of BER. Exploiting the property of the optimal caching placement which we derive,  the proposed algorithm has considerably reduced search space. It is shown that optimal caching placement is  not focusing on the file diversity gain only, contrary to \cite{5}; optimal caching placement is a balance between the file diversity gain and the channel diversity gain. 

\item We identify the conditions in terms of  the popularity factor, i.e., Zipf exponent, that the optimal placement can be found without running the proposed greedy algorithm, and derive the corresponding optimal caching placement in closed form. In particular, we derive  two special thresholds of popularity seeking an extreme of either file diversity gain or channel diversity gain, respectively. Furthermore, given enough proximity and transmit power, we show that optimal caching placement is parametrized solely with popularity. 
\end{itemize}

The rest of this paper is organized as follows: Section \ref{System_model} presents our system model. An optimization problem is formulated in Section \ref{Problem_formulation}. We propose the algorithm to find a solution in Section \ref{Optimal_caching_placement} and prove the optimality of the proposed algorithm. In 
Section \ref{Characterization}, we identify and analyze the tradeoff between the file diversity gain and the channel diversity gain. Numerical results are presented in Section \ref{Numberical Resutls}. Finally, Section \ref{Conclusion} concludes this paper.

\section{System Model}\label{System_model}
In this paper, we consider a cell composed of multiple clusters. 
In each cluster, the users can access $N$ helpers which are capable of
storing files in their memories. There is a file library whose size is $F$ which is strictly larger than $N$. i.e., $N<F$. Each helper can store one file from the library in advance, and the users
request a file in the library independently with probability $q_i$
according to the Zipf distribution with exponent $\gamma$, of which 
probability mass function is given by $q_i = \frac{\frac{1}{i^\gamma}}{\sum_{j=1}^F\frac{1}{j^\gamma}}.$
In each cluster, only one user is served for each file delivery period, which corresponds to orthogonal multiple access. Assuming orthogonal frequency allocation across clusters, inter-cluster interference is not considered in this paper, and thus we
consider one cluster in isolation for tractable analysis. In addition, for simplicity, we assume that based on open loop power control, all of the users in a cluster have the same average received signal-to-noise ratio (SNR) from helpers regardless of their positions, but they still suffer from independent small-scale fading effects. As a result, the multi-user
scenario is simplified into single-user one where a user who requests a file
is served at every moment.
\begin{figure}[t!]
    \centering
    \includegraphics[scale=0.27]{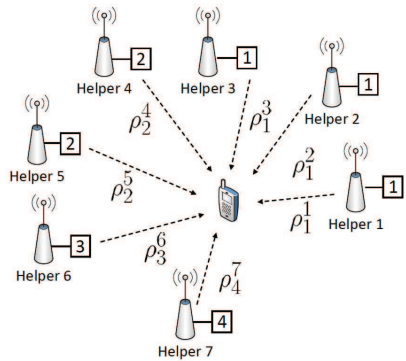}
    \caption{An example of caching placement in a cluster}
    \label{fig_system_model}
\end{figure}

If the requested file is already stored in any helper's memory in a
cluster, the requested file is transmitted from the helper which
cached the file. Throughout this paper, we call this type of
communication \textit{cluster communication}. If the requested
file is cached by more than one helper, a helper which has
the largest channel gain is selected by the requesting user, using
perfect channel state information. Fig. \ref{fig_system_model} shows an example of caching placement for a system with seven helpers, where the instantaneous received SNR from helper $k$ which caches the $i$th file is denoted by $\rho_i^k$; the average of $\rho_i^k$ is represented as $\bar{\rho}$. If the first file is requested by the user, since there are
three helpers which  has the file, the requesting user
selects the helper whose received SNR is the maximum among
the three helpers and informs the corresponding helper to
deliver the the first file. On the other hands, if the file requested by the user is not stored in the cluster, the BS serves the requesting user with average received SNR,
$\bar{\nu}$; namely, the file is transmitted
through \textit{cellular communication}. We denote the ratio between the average received SNRs of cluster and cellular communication as $\beta$ such that $\beta = \bar{\rho}/\bar{\nu}$, and 
	owing to closer distances to caching helpers than to the macro base station,  $\beta$ is reasonably assumed to be greater than one. However,  instantaneous channels gains from helpers and BS follow Rayleigh distribution. 

Packet error rate (PER) and frame error rate (FER) are practical performance metrics. However, it is hard to mathematically analyze the impacts of caching placement and channel fading with PER or FER because an expression of PER or FER incorporating all affecting factors, such as channel coding and upper layer parameters, is usually not available. 
	On the other hand, FER or PER is in general represented as a function of BER. Therefore, minimizing BER results in reducing FER or PER, albeit not linearly proportional. In this context, we study optimal caching placement to minimize BER, which will also proportionally reduce PER or FER. Assuming arbitrary coherent modulation/demodulation, the instantaneous BER in general form is given as \cite{gs} $p_e = \alpha_0 Q\left(\sqrt{\alpha_1 \lambda}\right)$, where $\lambda$ is the instantaneous received SNR, $\alpha_0$ , and $\alpha_1$ are appropriate constants for each type of modulation, and $Q\left(x\right)$ is Q-function defined as $Q\left(x\right) = \frac{1}{\sqrt{2\pi}} \int_x^{\infty} \exp\left(-\frac{u^2}{2}\right) du.$ For simplity, we assume $\alpha_0 = 1$ and $\alpha_1 = 2$, which  corresponds to QPSK, but generalization for arbitrary $\alpha_0$ and $\alpha_1$ is straightforward and the assumption of $\alpha_0 = 1$ and $\alpha_1 = 2$ does not lose any insights. Then, a file is transmitted over many coherent periods and the corresponding average BER of the file is obtained as
$
    \bar{p}_{e} =   \mathbb{E}_{\lambda}\left[Q\left(\sqrt{2\lambda}\right)\right]. 
$

\section{Problem Formulation}\label{Problem_formulation}
Because caching placement affects the average BER, in this section, we derive the average BER as a function of caching placement.  
Let us first derive the average BER of the $i$th file, denoted by $p_e\left(n_i\right)$, for given caching placement $\mathbf{n}^{N} = \left[n_1,\cdot,n_F\right]$ where $n_i$ is the number of helpers which cache the $i$th file such that $\sum_{i=1}^F n_i = N$. In the clustered system, since each helper provides equal average received SNR, caching a file in which helper does not matter in average sense, caching placement can be characterized with the number of files cached in the cluster. If there are no helpers that stored the $i$th file in its memory (i.e., $n_i=0$), the BS delivers the file to the user. In this case, since the average received SNR from BS is $\bar{\nu}$, the average BER with cellular communication,
$p_{e}^{\textrm{cellular}}$, is calculated as
\begin{align}
p_{e}^{\textrm{cellular}} &= \int_0^{\infty} Q(\sqrt{2x})\frac{1}{\bar{\nu}} \exp\left(-{\frac{x}{\bar{\nu}}}\right) dx = \frac{1}{2} \left( 1-
\sqrt{\frac{\bar{\nu}}{1+\bar{\nu}}}\right). \label{eq_c_error}
\end{align}
Otherwise, if $n_i\ne0$,  the $i{\textrm{th}}$ file 
is delivered via cluster communication. Since the helper with the largest channel gain among $n_i$ helpers is chosen
to deliver the $i{\textrm{th}}$ file to the user, the received SNR of cluster communication follows the distribution of
$\rho_{n_i}^{\max}=\max_k \rho_i^k$, of which PDF is
\begin{align}
f_{\rho_{n_i}^{\max}}(x)=\frac{n_i}{\bar{\rho}}\exp\left({-\frac{x}{\bar{\rho}}}\right)
\left(1-\exp\left({-\frac{x}{\bar{\rho}}}\right)\right)^{n_i -1}
.\label{eq_m_pdf}
\end{align} 
Using \eqref{eq_m_pdf}, the average BER of the $i$th file via  cluster
communication conditioned on $n_i$ cached helpers, $p_{e}^{\textrm{cluster}}(n_i)$
is obtained as 
	\begin{align}
	p_e^{\textrm{cluster}} \left(n_i\right) &= \int_0^{\infty} Q\left(\sqrt{2x}\right) f_{\rho_{n_i}^{\max}}\left(x\right) dx \label{eq_cl_error_1} \\
	&= \frac{1}{2} \sum_{m=0}^{n} {n\choose m}\left(-1\right)^m \sqrt{\frac{\bar{\rho}}{m+\bar{\rho}}},
	\label{eq_cl_error}
	\end{align}
Then, $p_{e}(n_i)$ is obtained as
\begin{align}\label{eq_error}
p_{e}(n_i) = \left\{ \begin{array}{lc}
p_{e}^{\textrm{cellular}}, &\mathrm{for}~  n_i=0,\\
p_{e}^{\textrm{cluster}}(n_i) & \mathrm{for}~ n_i > 0 \\
\end{array} \right. .
\end{align}
Since the $i{\textrm{th}}$ file is requested by the user with
probability $q_i$,  given caching placement $\mathbf{n}^N$, the average BER 
is derived as
\begin{align*}
\numberthis \bar{p}_e(\mathbf{n}^N) = \sum_{i=1}^{F} q_i \left[
\mathbf{1}\left(n_i\ne0\right)p_{e}^{\textrm{cluster}}(n_i) +
\mathbf{1}\left(n_i=0\right)p_{e}^{\textrm{cellular}}\right]\label{eq_avg_error},
\end{align*} where $\mathbf{1}(X)$ is the indication function; if event X is true, $\mathbf{1}(X)=1$ and otherwise, $\mathbf{1}(X)=0$.
Then, the optimal caching placement that minimizes the average BER
for the $N$-helper system can be obtained by solving the following optimization
problem:
\begin{align}
 \textbf{P}: \min_{\mathbf{n}^N} \bar{p}_e(\mathbf{n}^N) & \label{min_BER}\\
\nonumber \textrm{ subject to  } &\sum_{i=1}^F n_i = N, \hspace{10pt} n_i \in \mathbb{Z}^+,
\end{align} where $\mathbb{Z}^+$ is a set of non-negative integers.

\section{Optimal Caching Placement}\label{Optimal_caching_placement}
The design variable of \textbf{P} is an $N$-dimension integer vector $\mathbf{n}^N$. A problem which finds an optimal integer variable is called integer programming. In many cases, integer programming is too complicated to get a
closed-form solution. On that account, instead of solving the problem directly, we propose a greedy algorithm, which will be proven to be optimal under a mild condition.  Algorithm \ref{greedy_algorithm} presents the proposed greedy algorithm, where $\Delta p_e\left(n_k\right)$ is defined as
	\begin{align}
	\Delta p_e\left(n_k\right) = p_e\left(n_k\right) - p_e\left(n_k+1\right) = \left\lbrace \begin{array}{lc}
	p_e^{\textrm{cellular}} - p_e^{\textrm{cluster}}\left(1\right) & \textrm{for} \hspace{10pt} n_k = 0 \\
	p_e^{\textrm{cluster}}\left(n_k\right) - p_e^{\textrm{cluster}}\left(n_k+1\right)   & \textrm{for} \hspace{10pt} n_k \geq 1
	\end{array} \right. . \label{eq_delta_p_1}
	\end{align}
	By using \eqref{eq_c_error} and \eqref{eq_cl_error}, $\Delta p_e\left(n_k\right)$ is given as 
	\begin{align}
		\Delta p_e\left(n_k\right) = \left\lbrace \begin{array}{lc}
		\frac{1}{2}\left(\sqrt{\frac{\bar{\rho}}{1+\bar{\rho}}} - \sqrt{\frac{\bar{\rho}}{\beta + \bar{\rho}}}\right) & \textrm{for} \hspace{10pt} n_k = 0  \\
		\frac{1}{2} \sum_{m=0}^{n_k} {n_k \choose m} \left(-1\right)^m \sqrt{\frac{\bar{\rho}}{m+1+\bar{\rho}}}   & \textrm{for} \hspace{10pt} n_k \geq 1 \label{eq_del_p_n}
		\end{array} \right. .
	\end{align} 
	As shown in Algorithm \ref{greedy_algorithm}, at the $m$th iteration, the placement obtained by the proposed greedy algorithm provides the lowest average BER for the system with $m+1$ helpers. This is because the proposed greedy algorithm compares the
	amount of reduction in BER, $q_k \Delta
	p_{e}\left(\left(\mathbf{n}^{m}\right)_k\right)$, for
	each file and adds one file whose contribution is the largest to
	the memory in the cluster, where $\left(\mathbf{n}^{m}\right)_k$
	is the $k$th element of vector $\mathbf{n}^{m}$.
	\begin{algorithm*}
		\caption{The Greedy Algorithm}\label{greedy_algorithm}
		\begin{algorithmic}[1]
			\State \textbf{input} $N$,~$F$,~$\gamma$,~$\bar{\nu}$, and $\bar{\rho}$
			\State \textbf{initialize} $\mathbf{n}^1 = \left[1,0,\cdots,0\right]$			
			\For{$m =1,2,\cdots, N-1$}
			\State $j = \argmax_{k \in F}q_k  \Delta p_{e}\left(\left(\mathbf{n}^{m}\right)_k\right)$
			\State $\left(\mathbf{n}^{m+1}\right)_j \gets \left(\mathbf{n}^{m}\right)_j + 1 $
			\EndFor		
			\State \textbf{return} $\mathbf{n}^{N}$
		\end{algorithmic}
	\end{algorithm*}
In general, a greedy algorithm is not global optimal; however, we
show that the proposed greedy algorithm can find an optimal
solution. Through the following
lemma and theorem, we prove the optimality of the proposed greedy
algorithm.

\begin{lemma}\label{lemma_1}
	For $\beta \geq 2$, BER gain of $i$th file decreases as $n_i$ increases
	\begin{align}
	\Delta p_{e}\left(m\right) > \Delta p_{e}\left(n\right) \hspace{20pt} \textrm{for} \hspace{20pt} m < n. 
	\label{ineq_lemma_1}
	\end{align}
    \begin{IEEEproof}
    	Please refer to Appendix \ref{pf_lemma_1}.
    \end{IEEEproof}
\end{lemma}

\begin{theorem}[Optimality of the Proposed Greedy Algorithm]\label{thm_1}
    For $\beta \geq 2$, the caching placement obtained by the proposed greedy algorithm, $\mathbf{n}^{N}_{\textrm{greedy}}$, is optimal for the $N$-helper system.
    \begin{IEEEproof}
    Please refer to Appendix \ref{pf_theorem_1}.
    \end{IEEEproof}
\end{theorem}
The proposed greedy algorithm has low complexity, however, we can reduce the complexity further with the following proposition.

\begin{proposition}\label{proposition_1}
    For non-uniform popularity (i.e., $\gamma \neq 0$), the elements of optimal caching placement $\mathbf{n}_{\textrm{opt}}^N$ must satisfy $ n_1 \geq n_2 \geq \cdots \geq n_F \geq 0 $
    \begin{IEEEproof}
    	Please refer to Appendix \ref{pf_prop_1}.
    \end{IEEEproof}
\end{proposition}
Proposition \ref{proposition_1} implies that since a lower index file is
requested more frequently, the placement which stores
more files whose index is low can decrease the average BER more.
\begin{remark}
		It is shown in the proof of Theorem 1 that caching placement of the $m$th iteration is optimal for the system with $m+1$ helpers. When there exist $m+1$ helpers, caching the $m+2$th file is strictly suboptimal since this implies one of the files more popular than the $m+2$th file is not cached. Hence, $n_{m+2} = 0$ for optimal caching placement. Moreover, by Proposition 1, $n_{k} = 0$ for $k \geq m +2$. Therefore, in the $m$th iteration which determines optimal caching placement of a system with $m+1$ helpers, it is unnecessary to compare the BER gain of the files less popular than the $m+2$th popular file. Consequently, the number of comparisons in the $m$th iteration is reduced from the size of file library $F$ to $m$. 
		Because  $N-1$ iterations are required to find optimal caching placement of a system with $N$ helpers, the total number of comparisons decreases from $(N-1)F$ to $\sum_{i=m}^{N-1} m = \left(N-1\right)\left(N-2\right)/2$, which is independent of $F$.
	\end{remark}

\section{Characterization of Optimal Caching Placement}\label{Characterization}
     Theorem \ref{thm_1} ensures that the proposed greedy algorithm finds the optimal
caching placement, given system parameters (i.e., $N$,~$F$,~$\gamma$,~$\bar{\nu}$, and $\bar{\rho}$).
    In this section, we identify the conditions that  the optimal placement can be found without running the proposed greedy
    algorithm, and derive the corresponding optimal placement. Before analyzing further, we first define the two types of gains
which can be reaped by caching: \textit{file diversity gain} and
\textit{channel diversity gain}.
        \begin{definition}[File Diversity Gain]
        The amount of BER reduced by adding a new file which is not cached in any helper, is called file diversity gain.
	    \end{definition}	
      Changing communication from cellular to cluster reduces average BER by means of the proximity between the helpers and the user.
     Obviously, file diversity gain is proportional to the popularity of the file. Hence, the file diversity gain of the $k$th file is defined as
     	\begin{align}
     	 g_k^{\textrm{file}} = q_k \Delta p_e\left(0\right) \label{def_f_gain}
     	\end{align}
     \begin{definition}[Channel Diversity Gain]
        The amount of BER reduced by adding a file cached already by some helpers, is called channel diversity gain.
    \end{definition}

    As we consider the effect of wireless fading channel such as small-scale fading, if we increase the number of helpers which store a specific file, the selection pool of channel links enlarges and 
    BER of the file correspondingly improves due to the channel diversity. The channel diversity gain of the $k$th file is given as a function of the number of helpers that caches the $k$th file and the popularity of the $k$th file. The channel diversity gain of the $k$th popular file is written as
        	\begin{align}
        	g^{\textrm{channel}}_k\left(n_k\right) = q_k \Delta p_e\left(n_k\right) \hspace{10pt} \textrm{for} \hspace{10pt} n_k \geq 1.\label{def_ch_gain}
        	\end{align}
\subsection{Even caching placement}
In this subsection, we derive the condition when even
placement,
$\left[n_1,...,n_N,n_{N+1},...,n_F\right]=\left[1,...,1,0,...,0\right]$,
is optimal. In order that the output of the algorithm
becomes even placement, at each iteration, the file which was not
cached in the past iterations need to be cached. This happens when
the maximum file diversity gain, is larger than the maximum channel diversity gain,
for all iterations. On the other
hand, if there exists any single iteration at which there is a file
which has larger channel diversity gain than the maximum of file
diversity gain, even caching placement cannot be realized by the greedy algorithm. The
following proposition reveals the condition when even placement
is optimal.

    \begin{proposition}[Optimality of Even Caching Placement]\label{proposition_2}
        For $\beta \geq 2$, even caching placement is optimal if and only if $\gamma \leq \gamma_0$, where $\gamma_0$ is defined as
        \begin{align}
            \gamma_0 = \frac{1}{\log{N}} \left(\log{\left(1-\sqrt{\frac{1+\bar{\rho}}{\beta+\bar{\rho}}}\right)} - \log{\left(1-\sqrt{\frac{1+\bar{\rho}}{2+\bar{\rho}}}\right)}\right).
        \end{align}
    \end{proposition}
    \begin{IEEEproof}
    	Please refer to Appendix \ref{pf_prop_2}.
    \end{IEEEproof}
This proposition states that if user's preference 
for files is sufficiently unbiased (i.e., low Zipf exponent $\gamma$),
even caching placement is a reasonable approach toward optimal
performance.
    \begin{corollary}
        For any $\beta$, even caching placement cannot be optimal if $\gamma \ge \gamma_0^{\prime}$, where
        \begin{align}
            \gamma_0^{\prime} = -\frac{\log{\left(1-\sqrt{\frac{1+\bar{\rho}}{2+\bar{\rho}}}\right)}}{\log{N}}.\label{eq_gamma_0}
        \end{align}
    \end{corollary}
   \begin{IEEEproof} Since $\gamma_0$ is an increasing function of $\beta$ and is bounded, we have
        $
            \gamma_0  < \gamma_0^{\prime} = \lim_{\beta \rightarrow \infty} \gamma_0 
        $       
         which implies that if $\gamma \geq \gamma'_0$ then, $\gamma > \gamma_0$. By contraposition of Proposition 2, even caching placement cannot be optimal if $\gamma \geq \gamma'_0$.        
    \end{IEEEproof}
    This corollary implies that no matter how the received SNR of cluster communication is large, if popularity is highly biased, channel diversity gain should be considered. 
\subsection{Single-file caching placement}
    On the other extreme, if the smallest channel diversity gain is larger than the largest file diversity gain at every iteration, the output of the algorithm
    is single-file caching, i.e., $\left[n_1,n_2,...,n_F\right]=\left[N,0,...,0\right]$. 
    Similarly, we can obtain an optimality condition of single-file caching. 
    	\begin{proposition}[Optimality of Single-file Caching]\label{proposition_3}
    				For $\beta \geq 2$, single-file placement is optimal if and only if $\gamma \geq \gamma_1$, where $\gamma_1$ is defined as
    				\begin{align}
    				\gamma_1 = \frac{1}{\log{2}}\log{\left(\frac{\left(\frac{1}{\sqrt{1+\bar{\rho}}} - \frac{1}{\sqrt{\beta+\bar{\rho}}}\right)}{ \sum_{m=0}^{N-1} {N-1 \choose m} \left(-1\right)^m \sqrt{\frac{1}{m+1+\bar{\rho}}}}\right)}.
    				\end{align}
    			\end{proposition}
    \begin{IEEEproof}
    	Please refer to Appendix \ref{pf_prop_3}.
    \end{IEEEproof}
    \subsection{Doubly caching placement}
        When the Zipf exponent is between $\gamma_0$ and $\gamma_1$, owing to the following lemmas, we can analyze the structure of optimal placement in high SNR regime.
        \begin{lemma}\label{lemma_2}
            	As $\bar{\rho} \rightarrow \infty$, $g_k^\textrm{channel}\left(n_k\right) = o\left(g_j^\textrm{channel}\left(1\right)\right)$ and $g_k^\textrm{channel}\left(n_k\right) = o\left(g_j^\textrm{file}\right)$ for $n_k \geq 2$
        \end{lemma}
        \begin{IEEEproof}
        	Please refer to Appendix \ref{pf_lemma_2}.
        \end{IEEEproof}

Lemma \ref{lemma_2} reveals that the channel diversity gain of multiple helpers decreases much faster than both of channel diversity gain of one helper and file diversity gain, in high SNR regime. Hence, if one more helpers cache the file,  any file which was cached by a single helper or none of helpers has larger gain than any other files cached by multiple helpers, regardless of the popularity of files in high SNR regime.  

Now, we propose doubly caching placement with $k$, in which up to the $k$-th popular files  are stored in two helpers and the other files starting from the $k$+1-th popular file are stored in one helper until all helpers are occupied. In the following proposition, the proposed doubly caching placement is shown to be optimal for a certain range of the Zipf exponent.

        \begin{proposition}[Optimality  Condition on Doubly  Caching Placement with $k$]\label{proposition_4} As $\bar{\rho} \rightarrow \infty$, $\bar{\nu} \rightarrow \infty$, and $\frac{\bar{\rho}}{\bar{\nu}} \rightarrow \beta \geq 2$, doubly caching placement with $k < \left\lfloor \frac{N}{2} \right\rfloor$ is optimal if and only if $\gamma_2\left(k\right) \leq \gamma \leq \gamma_3\left(k\right)$ and doubly caching placement with $k = \left\lfloor \frac{N}{2} \right\rfloor$ is optimal if and only if $ \gamma \geq \gamma_2\left(\left\lfloor \frac{N}{2} \right\rfloor\right)$, where
            \begin{align}
                \gamma_2\left(k\right) &= \frac{1}{\log{\left(N-k+1\right)}-\log{k}}\left( \log{\left(1-\sqrt{\frac{1+\bar{\rho}}{\beta+\bar{\rho}}}\right)}-\log{\left(1-\sqrt{\frac{1+\bar{\rho}}{2+\bar{\rho}}}\right)}\right), \label{eq_gamma_2}\\
                \gamma_3\left(k\right) &= \frac{1}{\log{\left(N-k\right)}-\log{\left(k+1\right)}}\left( \log{\left(1-\sqrt{\frac{1+\bar{\rho}}{\beta+\bar{\rho}}}\right)}-\log{\left(1-\sqrt{\frac{1+\bar{\rho}}{2+\bar{\rho}}}\right)}\right).\label{eq_gamma_3}
            \end{align}
        \end{proposition}
        \begin{IEEEproof}
        	Please refer Appendix \ref{pf_prop_4}.
        \end{IEEEproof}

    With Propositions \ref{proposition_2}, \ref{proposition_3} and \ref{proposition_4},  optimal caching placement can be obtained without running the greedy algorithm
    by comparing the Zipf exponent of popularity, $\gamma$, with the thresholds, $\gamma_i$, $i=0,...,3$. That is, if we have information about the file preference of users, then we can determine the optimal caching placement of helpers.

    Until this point, we focused on the case when $\beta$ is larger than or equal to two. The condition $\beta \geq 2$ is necessary for file diversity gain of each file to be larger than channel diversity gain of the file from one to two helpers. The regime where $\beta \geq 2$ implies that the average received SNR from helpers is larger than that from the macro base station by 3 dB. Given the proximity of femto base stations, it is known that the average received SNR of a femto cell user  is approximately 10 dB higher than that of a macro cell user \cite{RB4}, so  $\beta \geq 2$ is typically achieved. We identified the optimal caching placement in the region of $\beta \geq 2$.
    
    For $\beta < 2$, fortunately, we can find optimal caching placement without running the proposed algorithm in high SNR for $\beta <2$.

    \begin{proposition}[Optimality of Doubly Caching Placement with $k =\left\lfloor\frac{N}{2}\right\rfloor$ for $\beta<2$]\label{proposition_5}
         As $\bar{\rho} \rightarrow \infty$, $\bar{\nu} \rightarrow \infty$, and $\frac{\bar{\rho}}{\bar{\nu}} \rightarrow \beta < 2$, doubly caching placement with $k=\lfloor\frac{N}{2}\rfloor$ is optimal. 
    \end{proposition}

    \begin{IEEEproof}
    	Please refer Appendix \ref{pf_prop_5}.
    \end{IEEEproof}   
   Proposition \ref{proposition_5} verifies the analytic insights;    smaller $\beta$ brings less file diversity gain since the gain of first caching is higher as the SNR difference between BS and helpers is larger. Consequently, when $\beta$ is less than $2$, channel diversity gain is more preferred. Combining this fact with Lemma \ref{lemma_2}, we can naturally conclude that caching all files in two helpers becomes optimal for $\beta < 2$ in high SNR.

    Table \ref{tbl_summary} summarizes the results of Section \ref{Characterization}; optimal caching placement depending on the range of Zipf exponent and $\beta$. The optimal caching placement in the region of $\beta \geq 2$ is identified. For $\beta <2$, the solution when SNR is not large  is unknown due to analytic intractability involved in Lemma 1 but  optimal caching placement is presented when SNR is high. 
     \begin{table}[!t]
     	\caption{Optimal Caching placement depending on the range of Zipf exponent}
     	\centering
     	\label{tbl_summary}
     	\begin{tabular}{|l|c|c|c|}
     		\hline
                                                                                          & In low and intermediate SNR                       & In  high SNR\\
     		\hline
     		$\beta\ge2$, $\gamma \leq \gamma_0$                                           & \multicolumn{2}{|c|}{Even caching placement}\\
     		\hline
            $\beta\ge2$, $\gamma_2\left(k\right) \leq \gamma \leq \gamma_3\left(k\right)$ & Output of the proposed algorithm & Doubly caching placement with $k$ \\
     		\hline
     		$\beta\ge2$, $\gamma \geq \gamma_1$                                             & \multicolumn{2}{|c|}{Single caching placement}\\
     		\hline
     		$\beta<2$                                                                     & Unknown & Doubly caching placement with $k=\lfloor\frac{N}{2}\rfloor$ \\
     		\hline
     	\end{tabular}
     \end{table}
\section{Caching Multiple Files in Helpers}
	In this section, we extend the proposed greedy algorithm to when each helper caches up to $M$ files. When each helper can cache up to $M$ files, the optimization of minimizing the BER requires additional constraint $n_i \leq N$. Since caching the same file at a helper more than twice is strictly not optimal, $n_i \leq N$  is additionally imposed to avoid unnecessary repetition.
	 Note that the total memory constraint $\sum_{i=1}^F n_i = N$ in \textbf{P} is replaced by $\sum_{i=1}^F n_i = NM$ which is the numbers of files when $N$ helpers can cache when each helper stores $M$ files.
	To solve this problem, we propose a modified greedy algorithm, namely \textit{$M$-round greedy algorithm} which consists of $M$-repetitions of the original greedy algorithm proposed for single file caching. Since the original greedy algorithm is proven to be optimal when each helper can cache a single file, we fill the memory of each helper in each round by using the original greedy algorithm. However, in the original greedy algorithm $n_i \leq N$ is not addressed. Therefore, to cover this constraint, if $n_i = N$ is satisfied at a certain round, the $i$th file is discarded after that round until the algorithm finishes.  
	\begin{figure}[t!]
		\centering
		\includegraphics[scale=0.7]{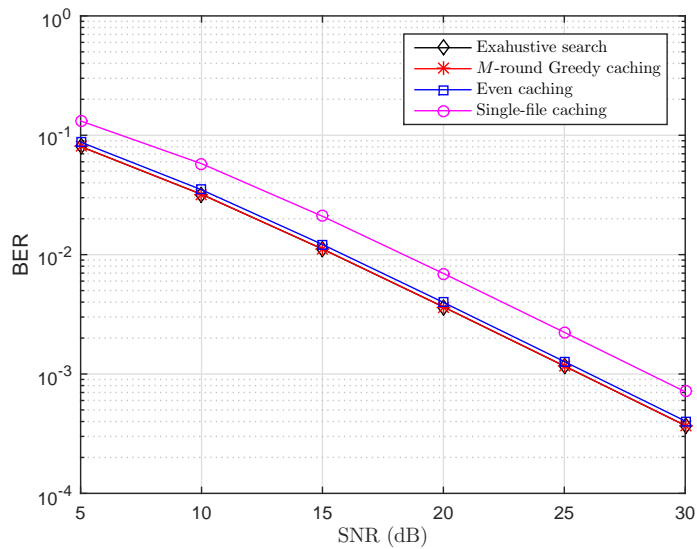}
		\caption{Performance of $M$-round greedy algorithm for $N = 5, M = 5, F=50$, and $\gamma = 0.6$} \label{fig_sim_M_greedy}
	\end{figure}	
	Although the proposed $M$-round greedy algorithm is suboptimal, Fig. \ref{fig_sim_M_greedy} shows that its performance is close to the BER achieved by the optimal one found by exhaustive searches.

\section{Numerical Results}\label{Numberical Resutls}

In this section, we demonstrate some numerical results that verify our analysis. We assume  $F = 20, N = 10, \gamma = 0.6,$ and $\beta = 5$ [dB] as a default setting \cite{1}. Depending independent variables, the simulation environment is slightly changed and mentioned in each subsection.

\subsection{Optimality of the proposed greedy algorithm}
 In this subsection, we verify that the proposed greedy algorithm finds the optimal caching placement. Table \ref{tbl_0.6} shows optimal caching placement for various numbers of helpers for $\bar{\rho} = 15$ [dB], which is found by numerical full search. It is exhibited that optimal caching placement changes in a greedy way as the number of helpers increases. That is, only one component of optimal caching placement is changed as $N$ increases to $N+1$.

\begin{table}[b!]
  \caption{Optimal caching placement for different number of helpers}
    \centering
    \label{tbl_0.6}
    \begin{tabular}{|c|c|c|c|}
       \hline
        \bfseries Number of helpers & \bfseries Optimal caching placement & \bfseries Number of helpers & \bfseries Optimal caching placement\\
        \hline
        $N=1$ & $\left[1,0,0,0,0,0,0,0,\cdots,0\right]$ & $N=6$ & $\left[2,2,1,1,0,0,0,0,\cdots,0\right]$ \\
        \hline
        $N=2$ & $\left[1,1,0,0,0,0,0,0,\cdots,0\right]$ & $N=7$ & $\left[2,2,1,1,1,0,0,0,\cdots,0\right]$ \\
        \hline
        $N=3$ & $\left[2,1,0,0,0,0,0,0,\cdots,0\right]$ & $N=8$ & $\left[2,2,1,1,1,1,0,0,\cdots,0\right]$\\
        \hline
        $N=4$ & $\left[2,1,1,0,0,0,0,0,\cdots,0\right]$ & $N=9$ & $\left[2,2,2,1,1,1,0,0,\cdots,0\right]$\\
        \hline
        $N=5$ & $\left[2,1,1,1,0,0,0,0,\cdots,0\right]$ & $N=10$ & $\left[2,2,2,1,1,1,1,0,\cdots,0\right]$\\
        \hline
    \end{tabular}
\end{table}

\subsection{Optimal caching placement in high SNR region}
Table \ref{tbl_highSNR} exhibits optimal caching placement when  $\bar{\rho} = 5$ [dB] and $\bar{\rho} = 40$ [dB]. In low SNR region (i.e., $\bar{\rho} = 5$ [dB]), the optimal caching placement is highly biased as the Zipf exponent grows because it is beneficial to offer  robust links for top-ranked files. However, in high SNR region (i.e., $\bar{\rho} = 40$ [dB]), the optimal caching placement follows  Proposition \ref{proposition_4};  from \eqref{eq_gamma_2} and \eqref{eq_gamma_3}, we have $\gamma_2(3)=0.78$, $\gamma_3(3)=1.38$ and $\gamma_2(4)=1.38$, $\gamma_3(4)=4.23$, and thus the proposed doubly caching placement with $k=3$ (or $k=4$) becomes optimal when
the Zipf exponent is in $\left[0.78,~1.38\right]$ (or $\left[1.38,~4.23\right]$).

\begin{table}[!t]
    \caption{Optimal Caching placement of low and high SNR for different Zipf exponent}
    \centering
    \label{tbl_highSNR}
    \begin{tabular}{|c|c|c|}
        \hline
        Zipf exponent & $\bar{\rho}=5$ [dB] & $\bar{\rho}=40$ [dB]\\
        \hline
        $\gamma = 1$ & $\left[2,2,2,1,1,1,1,0,\ldots,0\right]$ & $\left[2,2,2,1,1,1,1,0,\ldots,0\right]$\\
        \hline
        $\gamma = 2$ & $\left[4,2,2,1,1,0,0,0,\ldots,0\right]$ & $\left[2,2,2,2,1,1,0,0,\ldots,0\right]$\\
        \hline
        $\gamma = 3$ & $\left[5,3,2,0,0,0,0,0,\ldots,0\right]$ & $\left[2,2,2,2,1,1,0,0,\ldots,0\right]$\\
        \hline
        $\gamma = 4$ & $\left[6,3,1,0,0,0,0,0,\ldots,0\right]$ & $\left[2,2,2,2,1,1,0,0,\ldots,0\right]$\\
        \hline
        $\gamma = 5$ & $\left[7,3,0,0,0,0,0,0,\ldots,0\right]$ & $\left[2,2,2,2,2,0,0,0,\ldots,0\right]$\\
        \hline
    \end{tabular}
\end{table}

\subsection{Performance comparison with other caching placement}
We compare BER performance of optimal caching placement with other caching placement for various system environments. 
\begin{figure}[!t]
    \centering
    \includegraphics[scale=0.7]{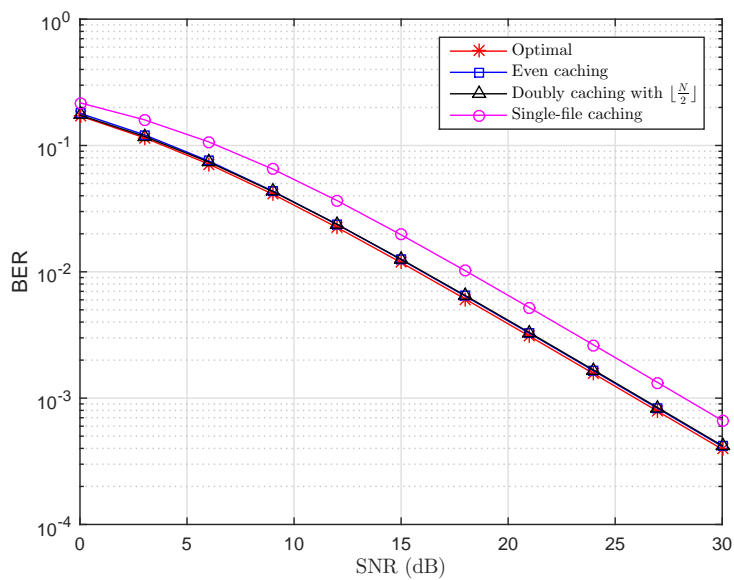}
    \caption{Average BER versus received SNR of cluster communication for $\gamma=0.6$.}
    \label{fig_sim1}
\end{figure}
In Fig. \ref{fig_sim1}, the average BERs of different caching placement are shown for various SNR of cluster communication under the default system parameters. This figure reveals that doubly caching placement achieves almost the same BER performance as optimal caching placement in all SNR region. Although even caching placement is known to be optimal without consideration of wireless aspects \cite{5}, it is rather outperformed by doubly caching placement in all SNR region. 

\begin{figure}[!t]
    \centering
    \includegraphics[scale=0.70]{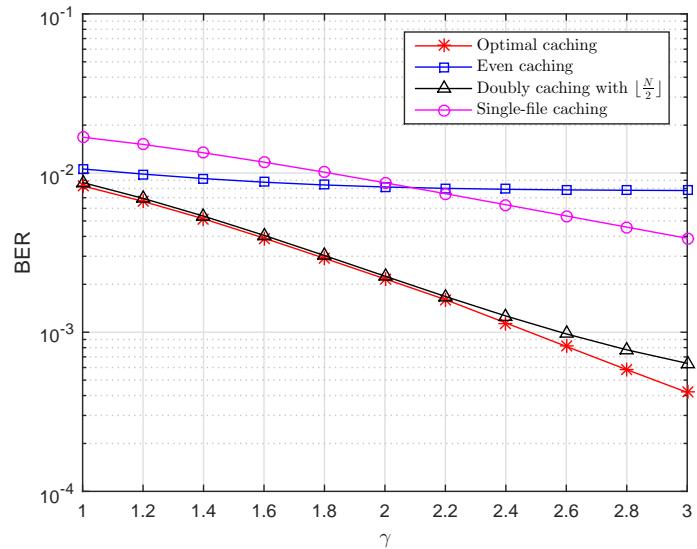}
    \caption{Average BER versus $\gamma$ for SNR$=15$ [dB].}
    \label{fig_sim2}
\end{figure}

The effects of the Zipf exponent is considered in Fig. \ref{fig_sim2} when $\bar{\rho}=15$ [dB]. As the Zipf exponent increases, the frequency of requesting the most popular file increases and thus the BER of high-ranked file dominates the average BER. Consequently, single-file caching placement shows better performance than even caching placement in high $\gamma$, but the opposite result is observed in low $\gamma$. On the other hand, doubly caching placement still show comparable BER performance with optimal caching placement for all values of the Zipf exponent $\gamma$.

\section{Conclusion}\label{Conclusion}
We proposed an optimal greedy algorithm for caching placement in wireless femto-caching network. The proposed algorithm minimizes the average bit error rate with low computational complexity, exploiting the property of optimal caching placement which we derived.  We also identified and explored the tradeoff between file diversity gain and channel diversity gain to minimize the average bit error rate. We derived  two special thresholds of popularity seeking an extreme of either file diversity gain or channel diversity gain, respectively, which provide a useful caching placement guideline without running the proposed algorithm. Furthermore, given enough proximity and transmit power, we showed that optimal caching placement was parametrized solely with popularity. 


%





\appendices
\def\thesection{\Alph{section}}%
\def\thesectiondis{\Alph{section}}%
\section{Proof of Lemma \ref{lemma_1}}\label{pf_lemma_1}
\setcounter{equation}{0}
\renewcommand{\theequation}{A.\arabic{equation}}
		We prove this lemma according to the range of $n$: $n=1$ and $n > 1$. For $n=1$, since zero is only integer which satisfies inequality $m < 1$, \eqref{ineq_lemma_1} becomes $\Delta p_{e}\left( 0\right) > \Delta p_{e}\left(1 \right).$
		Using \eqref{eq_del_p_n}, $\Delta p_e\left(0\right)$ and $\Delta p_e\left(1\right)$ become
		\begin{align}
		\Delta p_e\left(0\right) &= \frac{1}{2} \left( \sqrt{\frac{\bar{\rho}}{1+\bar{\rho}}} - \sqrt{\frac{\bar{\rho}}{\beta+\bar{\rho}}}\right), \label{eq_del_p_0} \\
		\Delta p_e\left(1\right) &= \frac{1}{2} \left( \sqrt{\frac{\bar{\rho}}{1+\bar{\rho}}} - \sqrt{\frac{\bar{\rho}}{2+\bar{\rho}}}\right).\label{eq_del_p_1}
		\end{align}
		Obviously, we have, for $\beta \geq 2$,
		\begin{align}
		\Delta p_e\left(0\right) - \Delta p_e\left(1\right) = \frac{1}{2} \left( \sqrt{\frac{\bar{\rho}}{2+\bar{\rho}}} - \sqrt{\frac{\bar{\rho}}{\beta+\bar{\rho}}}\right) \geq 0.
		\end{align}
		Second, for the case when $n>1$, we prove the equivalent statement that $\Delta p_e\left(n\right) >\Delta p_e\left(n+1\right)$ for $\forall n>1$.  The equivalence is established by the two facts;  (1)  $\forall m < n$ and $n>1$, $\Delta p_e\left(m\right) >\Delta p_e\left(n\right)$ obviously implies  $\Delta p_e\left(n\right) >\Delta p_e\left(n+1\right)$. (2) Conversely,  $\Delta p_e\left(n\right) > \Delta p_e\left(n+1\right)$ can be extended to $\Delta p_e\left(n-1\right) >\Delta p_e\left(n\right)$, and using mathematical induction, we can conclude $\Delta p_e\left(m\right) > p_e\left(n\right)$ for $\forall m$ such that $m<n$.
		
		Now, we prove the equivalent statement. Expressing the $\Delta p_e\left(n\right)$ with \eqref{eq_cl_error_1},
		\begin{align}
			\Delta p_e\left(n\right) = \int_0^{\infty} Q\left(\sqrt{2x}\right)\left(f_{\rho^{\max}_{n}}\left(x\right)- f_{\rho^{\max}_{n+1}}\left(x\right)\right) dx. \label{eq_del_p_n_Q}
		\end{align}
		$\Delta p_e\left(n\right) - \Delta p_e\left(n+1\right)$ can be rewritten as 
		\begin{align}
		\Delta p_e\left(n\right) - \Delta p_e\left(n+1\right) &= \int_0^{\infty} Q\left(\sqrt{2x}\right)\left(f_{\rho^{\max}_{n}}\left(x\right)- 2f_{\rho^{\max}_{n+1}}\left(x\right) + f_{\rho^{\max}_{n+2}}\left(x\right)\right) dx \\
		& = \int_0^{\infty} Q\left(\sqrt{2x}\right)R\left(x,n \right) dx,
		\end{align} where $R\left(x,n\right)$ is defined as $R\left(x,n\right) = f_{\rho^{\max}_{n}}\left(x\right)- 2f_{\rho^{\max}_{n+1}}\left(x\right) + f_{\rho^{\max}_{n+2}}\left(x\right)$.
		Since $Q(x)$ has a positive value for all $x$, it is enough to show that
		\begin{align}
		R\left( x,n\right) >0 \hspace{10pt} \forall x \geq 0 \textrm{~and~} \forall
		n\ge1.
		\end{align}
		Using the given PDF formula of $f_{\rho^{\max}_{n}}\left(x\right)$ in \eqref{eq_m_pdf}, $R\left(x,n\right)$ becomes
		\begin{align}
		R\left(x,n\right) &=\frac{1}{\bar{\rho}}\exp\left(-\frac{x}{\bar{\rho}}\right) \left(1-\exp\left(-\frac{x}{\bar{\rho}}\right)\right)^{n-1} \\
		&\times \left[n - 2\left(n +1\right)\left(1-\exp\left(-\frac{x}{\bar{\rho}}\right)\right) + \left(n+2\right)\left(1-\exp\left(-\frac{x}{\bar{\rho}}\right)\right)^2\right]  \\
		& = \frac{n+2}{\bar{\rho}} \exp\left(-\frac{3x}{\bar{\rho}}\right)\left(1-\exp\left(-\frac{x}{\bar{\rho}}\right)\right)^{n-1}, 
		\end{align} which is greater than zero for arbitrary $x \geq
		0$ and $n \ge1$.

\section{Proof of Theorem \ref{thm_1}}\label{pf_theorem_1}
\setcounter{equation}{0}
\renewcommand{\theequation}{B.\arabic{equation}}
         This theorem is proved by induction. First, for $N=1$, we can readily show that $\mathbf{n}^{1}_{\textrm{opt}}=\mathbf{n}^{1}_{\textrm{greedy}}$.
          Supposing $\mathbf{n}^{N-1}_{\textrm{opt}}=\mathbf{n}^{N-1}_{\textrm{greedy}}$,
          we will show that $\mathbf{n}^{N}_{\textrm{opt}}=\mathbf{n}^{N}_{\textrm{greedy}}$ by contradiction.
          Suppose $\mathbf{n}^{N}_{\textrm{greedy}}$ is not optimal and there exists optimal caching placement
          for the  $N$-helper system such that
        \begin{align}
        \bar{p}_{e} \left(\mathbf{n}^{N}_{\textrm{opt}} \right) < \bar{p}_e \left(\mathbf{n}^{N}_{\textrm{greedy}}
        \right).
        \label{ineq_ber_N}
        \end{align}
        Following the Algorithm \ref{greedy_algorithm}, the average BER of $\mathbf{n}^{N}_{\textrm{greedy}}$ is given as
        \begin{align}
        \bar{p}_e \left(\mathbf{n}^{N}_{\textrm{greedy}} \right)
        &= \bar{p}_e(\mathbf{n}_{\textrm{opt}}^{N-1}) - \max_{k \in F} q_k \Delta p_{e}\left(\left(\mathbf{n}_{\textrm{opt}}^{N-1}\right)_k\right)
        .
        \label{eq_ber_opt}
        \end{align}
        Hence, substituting (\ref{eq_ber_opt}) into (\ref{ineq_ber_N}), we obtain the following inequality
        \begin{align}
        \bar{p}_{e} \left(\mathbf{n}^{N-1}_{\textrm{opt}} \right) - \bar{p}_e(\mathbf{n}_{\textrm{opt}}^{N}) > \max_{k \in F} q_k\Delta p_{e}\left(\left(\mathbf{n}_{\textrm{opt}}^{N-1}\right)_k\right)
        .
        \label{ineq_max}
        \end{align}
        Now, let us consider another caching placement $\mathbf{\tilde{{n}}}^{N-1}$ for the $N-1$ helpers system, of which elements are the same as $\mathbf{n}^N_{\textrm{opt}}$ except that one element which least increases
        the average BER is reduced by one. In other words, $\mathbf{\tilde{{n}}}^{N-1}$ comes from $\mathbf{n}^N_{\textrm{opt}}$ in a reverse-greedy way.
        \begin{align}
        \left(\mathbf{\tilde{n}}^{N-1}\right)_i = \left\lbrace \begin{array}{ccr}\left(\mathbf{n}_{\textrm{opt}}^{N}\right)_i~~~~\qquad \textrm{if~} i \neq j& \\
        \left(\mathbf{n}_{\textrm{opt}}^{N}\right)_i-1 \qquad\textrm{if~} i =
        j&\end{array}\right.\label{eq_c_placement_N-1},
        \end{align} where $j$ is defined as $j = \arg\min_{k \in F} q_k \Delta p_{e} \left(\left(\mathbf{n}^{N}_{\textrm{opt}}\right)_k-1\right)$.

        We note that the average BER of $\mathbf{\tilde{n}}^{N-1}$ must not be less than that of $\mathbf{n}^{N-1}_{\textrm{opt}}$; hence, we have
        \begin{align}
        \bar{p}_{e} \left(\mathbf{n}^{N-1}_{\textrm{opt}} \right) \leq \bar{p}_e \left(\mathbf{\tilde{n}}^{N-1} \right).
        \label{ineq_ber_N-1}
        \end{align}
        In addition, from the definition of $\mathbf{\tilde{n}^{N-1}}$,
        \begin{align}
        \bar{p}_e \left(\mathbf{\tilde{n}}^{N-1} \right) = \bar{p}_e \left (\mathbf{n}^{N}_{\textrm{opt}} \right) + \min_{k} q_k \Delta p_{e}\left(\left(\mathbf{n}^{N}_{\textrm{opt}}\right)_k-1\right)
        .
        \label{eq_ber_tilde}
        \end{align}
        By substituting (\ref{eq_ber_tilde}) into (\ref{ineq_ber_N-1}), (\ref{ineq_ber_N-1}) becomes
        \begin{align}
        \bar{p}_{e} \left(\mathbf{n}^{N-1}_{\textrm{opt}} \right) - \bar{p}_e \left (\mathbf{n}^{N}_{\textrm{opt}} \right) \leq \min_{k}  q_k\Delta p_{e}\left(\left(\mathbf{n}^{N}_{\textrm{opt}}\right)_k-1\right)
        .
        \label{ineq_min}
        \end{align}
        If we combine (\ref{ineq_max}) and (\ref{ineq_min}), we have the following inequality:
        \begin{align}
        \max_{k} q_k \Delta p_{e}\left(\left(\mathbf{n}^{N-1}_{\textrm{opt}}\right)_k\right) < \min_k q_k \Delta
        p_{e}\left(\left(\mathbf{n}^{N}_{\textrm{opt}}\right)_k-1\right).
        \label{ineq_max_min}
        \end{align}
        Now, to complete our proof by contradiction, we prove that \eqref{ineq_max_min}
        cannot be satisfied.
        Since $\sum_k \left(\mathbf{n}^{N}_{\textrm{opt}}\right)_k = N$ and $\sum_k \left(\mathbf{n}^{N-1}_{\textrm{opt}}\right)_k = N-1$, there exists at least one file index, $l\in\{1,...,F\}$ such that $    \left(\mathbf{n}^{N}_{\textrm{opt}}\right)_l \geq \left(\mathbf{n}^{N-1}_{\textrm{opt}}\right)_l +1$.
        According to Lemma \ref{lemma_1} and $    \left(\mathbf{n}^{N}_{\textrm{opt}}\right)_l \geq \left(\mathbf{n}^{N-1}_{\textrm{opt}}\right)_l +1$, the following inequality holds
        \begin{align}
        q_l \Delta p_{e}\left(\left(\mathbf{n}^{N}_{\textrm{opt}}\right)_l-1\right) \leq q_l\Delta
        p_{e}\left(\left(\mathbf{n}^{N-1}_{\textrm{opt}}\right)_l\right). \label{ineq_thm_1}
        \end{align}
        Since, $\min_k q_k \Delta
        p_{e}\left(\left(\mathbf{n}^{N}_{\textrm{opt}}\right)_k-1\right)$ is smaller than the left-hand side of \eqref{ineq_thm_1}, and \\
         $\max_{k} q_k \Delta p_{e}\left(\left(\mathbf{n}^{N-1}_{\textrm{opt}}\right)_k\right)$ is greater than the right-hand side of \eqref{ineq_thm_1}, we have
        \begin{align}
        \min_k q_k \Delta p_{e}\left(\left(\mathbf{n}^{N}_{\textrm{opt}}\right)_k-1\right) \leq \max_{k} q_k \Delta
        p_{e}\left(\left(\mathbf{n}^{N-1}_{\textrm{opt}}\right)_k\right),
        \end{align}
        which contradicts (\ref{ineq_max_min}).
        Consequently, there cannot exist another optimal caching placement which is not $\mathbf{n}^{N}_{\textrm{greedy}}$. This completes the proof.        
\section{Proof of Proposition \ref{proposition_1}}\label{pf_prop_1}
\setcounter{equation}{0}
\renewcommand{\theequation}{C.\arabic{equation}}
		By contradiction, we prove Proposition \ref{proposition_1}. Suppose the optimal caching placement for the $N$-helper system, $\mathbf{n}_{\textrm{opt}}^N$,
		does not satisfy the Proposition \ref{proposition_1}. Then, there exists at least one $l\in\{1,...,F\}$ which
		satisfies 
		\begin{align}
		(\mathbf{n}^N_\textrm{opt})_l < (\mathbf{n}^N_\textrm{opt})_{l+1}. \label{ineq_rever_order}
		\end{align} 
		Consider another caching placement $\mathbf{\hat{n}}^{N}$ that has the same element with $\mathbf{n}_{\textrm{opt}}^N$ except that $n_l$ and $n_{l+1}$ are switched with each other as follows:
		\begin{align}
		\left(\mathbf{\hat{n}}^{N}\right)_i = \left\{ \begin{array}{l}
		\left(\mathbf{n}_{\textrm{opt}}^{N}\right)_i~\qquad~~\mathrm{if}~  i \neq l \textrm{~and~} i \neq l+1\\
		\left(\mathbf{n}_{\textrm{opt}}^{N}\right)_{l+1}\qquad\mathrm{if}~  i=l\\
		\left(\mathbf{n}_{\textrm{opt}}^{N}\right)_{l}~~~\qquad\mathrm{if}~  i=l+1\\
		\end{array} \right. .
		\end{align}
		Then, the difference of average BER between two caching placement schemes becomes
		\begin{align}
			\bar{p}_e\left(\mathbf{n}_{\textrm{opt}}^N \right) - \bar{p}_e\left(\mathbf{\hat{n}}^{N} \right) = & q_l p_{e}\left(\left(\mathbf{n}_{\textrm{opt}}^N \right)_l\right) + q_{l+1} p_{e}\left(\left(\mathbf{n}_{\textrm{opt}}^N \right)_{l+1}\right) - q_l p_{e}\left(\left(\mathbf{\hat{n}}^{N} \right)_{l}\right) - q_{l+1}
			p_{e}\left(\left(\mathbf{\hat{n}}^{N} \right)_{l+1}\right)\nonumber\\
			=& \left(q_l - q_{l+1}\right) \left( p_{e}\left(\left(\mathbf{n}_{\textrm{opt}}^N \right)_l\right) - p_{e} \left(\left(\mathbf{n}_{\textrm{opt}}^N \right)_{l+1}\right) \right).\label{eq_non_nega}
			\end{align}
			Since the $l$-th popular file is more probable to be requested than the $l+1$-th popular file, the first term on the right-hand side on \eqref{eq_non_nega} is positive. \eqref{ineq_rever_order} indicates that we have more helpers which cache the $l+1$-th popular file than helpers which cache the $l$-th popular file. Consequently, the average BER of the $l+1$-th popular file is lower than that of  the $l$-th popular file because more helpers provide higher channel diversity gain which reduces BER.
	    Therefore, \eqref{eq_non_nega} is positive; we have
		$\bar{p}_e\left(\mathbf{n}_{\textrm{opt}}^N \right) > \bar{p}_e\left(\mathbf{\hat{n}}^{N} \right).$
		This contradicts to optimal caching placement.	
\section{Proof of Proposition \ref{proposition_2}}\label{pf_prop_2}
\setcounter{equation}{0}
\renewcommand{\theequation}{D.\arabic{equation}}
		First, we prove the necessary condition for optimality of even caching placement. The condition $\gamma \leq \gamma_0$ can be rewritten as 
			\begin{align}
			& \hspace{30pt}\gamma \leq \frac{1}{\log N} \left(\log \left( 1- \sqrt{\frac{1+\bar{\rho}}{\beta+\bar{\rho}}} \right) - \log \left( 1- \sqrt{\frac{1+\bar{\rho}}{2+\bar{\rho}}} \right)\right)\\
			&\iff \frac{N^{-\gamma}}{\sum_{i \in F} i^{-\gamma}} \left(1-\sqrt{\frac{1+\bar{\rho}}{\beta+\bar{\rho}}}\right) \geq \frac{1}{\sum_{i \in F} i^{-\gamma}} \left(1-\sqrt{\frac{1+\bar{\rho}}{2+\bar{\rho}}}\right)  \\
			&\iff g_N^{\textrm{file}} \geq g_1^{\textrm{channel}}(1). \label{ineq_gamma_0}
			\end{align}	
		 Note that \eqref{ineq_gamma_0} means that the file diversity gain of the $N$-th file (i.e., the least popular file among possibly cached files) is larger than the channel diversity gain of the most popular file cached in a single helper. 
		 In addition to that, by the order of popularity, both of file and channel diversity gain decreases as popularity of file becomes lower. Hence, for any $i\leq N$, $g_i^{\textrm{file}} \geq g_N^{\textrm{file}}$. Similarly, for any $ k \geq 1$, $g_1^{\textrm{channel}}\left(n\right) \geq g_k^{\textrm{channel}}\left(n\right)$. \\
		 Furthermore, by Lemma 1 which states that $\Delta p_e\left(m\right) > \Delta p_e\left(n\right)$ for $ m < n$, we can conclude $\Delta p_e\left(1\right) \geq \Delta p_e\left(n\right)$ for $n \geq 1$. Since $g_k^{\textrm{channel}}\left(n_k\right)$ is a product of popularity and $\Delta p_e\left(n_k\right)$, we have $g_1^{\textrm{channel}}\left(1\right) \geq g_k^{\textrm{channel}}\left(n_k\right)$ for $k \geq 1$ and $n_k \geq 1$. Combining this result with \eqref{ineq_gamma_0}, we have the following inequality:
		 	$
		 	g_i^{\textrm{file}} \geq g_N^{\textrm{file}} \geq g_1^{\textrm{channel}}\left(1\right) \geq g_k^{\textrm{channel}}\left(n_k\right)$ for $i<N$ which implies that lower bound of file diversity gain is larger than upper bound of channel diversity gain. As a consequence, file diversity gain is always larger than the maximum of channel diversity gain, which concludes optimality of even caching placement.\\		
		For the proof of converse, we will show the contraposition of the converse is true. The contraposition of the converse is that if $\gamma > \gamma_0$, even caching placement is not optimal.
		Similarly, $\gamma > \gamma_0$ implies
		$
		g^{\textrm{file}}_N < g^{\textrm{channel}}_1\left(1\right).
		$
		Then, obviously, before the $N$-th popular file is cached, the most popular file will be cached in two helpers. As a consequence, even caching placement cannot be optimal.	
\section{Proof of Proposition \ref{proposition_3}}\label{pf_prop_3}
\setcounter{equation}{0}
\renewcommand{\theequation}{E.\arabic{equation}}
	We first prove the sufficient condition for optimality: if $\gamma \geq \gamma_1$, then single-file placement is optimal.
	The condition $\gamma \geq \gamma_1$ can be rewritten as
	\begin{align}
	&\hspace{30pt} \gamma \geq \frac{1}{\log 2} \log\left( \frac{\left(\frac{1}{\sqrt{1+\bar{\rho}}} - \frac{1}{\sqrt{\beta+\bar{\rho}}}\right)}{\sum_{m=0}^{N-1} {N-1\choose m} \left(-1\right)^m \sqrt{\frac{1}{m+1+\bar{\rho}}}} \right) \\			
	&\iff \frac{1}{\sum_{i\in F} i^{-\gamma}} \frac{1}{2} \sum_{m=0}^{N-1} {N-1\choose m} \left(-1\right)^m \sqrt{\frac{\bar{\rho}}{m+1+\bar{\rho}}} \geq \frac{2^{-\gamma}}{\sum_{i\in F} i^{-\gamma}} \frac{1}{2} \left(\frac{1}{\sqrt{1+\bar{\rho}}} - \frac{1}{\sqrt{\beta+\bar{\rho}}}\right)  \\			
	&\iff g_1^{\textrm{channel}}\left(N-1\right) \geq g_2 ^{\textrm{file}}
	\end{align} 
	By Lemma \ref{lemma_1}, 
	$
	g_1^{\textrm{channel}}\left(n\right) \geq g_1^{\textrm{channel}}\left(N-1\right) \geq g_2^{\textrm{file}}$ for $n \leq  N-1
	$
	which implies that if $\gamma \geq \gamma_1$, channel diversity gain of the most popular file is always larger than file diversity gain of the second-popular file until $N$ helpers cache the most popular file. Thus, the maximum channel diversity gain is always larger than maximum of file diversity gain for all iterations and consequently single-file placement becomes optimal.	
	For proving the necessary condition, we prove the contraposition: if $\gamma < \gamma_1$, then single-file placement is not optimal.
	$\gamma < \gamma_1$ is equivalent to
	$
	g_1^{\textrm{channel}} \left(N-1\right) < g_2^{\textrm{file}}
	$
	which means the second-popular file will be cached before caching the first-popular one $N$ times. Hence, single-file placement cannot be optimal.
\section{Proof of Lemma \ref{lemma_2}}\label{pf_lemma_2}
\setcounter{equation}{0}
\renewcommand{\theequation}{F.\arabic{equation}}
	Based on \eqref{eq_del_p_n_Q}, after simple manipulation, channel diversity gain can be expressed as integral of Beta function. Using the two properties of the Beta function, $
	B\left(x+1,y\right) = \frac{x}{x+y} B\left(x,y\right)$ and $B\left(1,1+\frac{\bar{\rho}}{\sin^2\theta}\right)=\frac{1}{(2+\frac{\bar{\rho}}{\sin^2\theta})(1+\frac{\bar{\rho}}{\sin^2\theta})}$,
	we have, for any $n_l\ge1$,
	\begin{align}
	g_l^\textrm{channel}\left(n_l\right)
	& =  \frac{q_l}{\pi}\int_0^{\frac{\pi}{2}} \frac{\bar{\rho}}{\sin^2\theta} \frac{n_l}{n_l+1+\frac{\bar{\rho}}{\sin^2\theta}}B\left(n_l,1+\frac{\bar{\rho}}{\sin^2\theta}\right) d\theta\\
	&=\bar{\rho}^{-n_l} \frac{q_l}{\pi} \int_0^{\frac{\pi}{2}} \frac{ G\left(n_l,\sin^2\theta,\bar{\rho}\right)}{\sin^2\theta} d\theta, \label{eq_ch_gain_1}
	\end{align}
	where $G\left(n_l,\sin^2\theta,\bar{\rho}\right)$ is defined as
	\begin{align}
	G\left(n_l,\sin^2\theta,\bar{\rho}\right) = \frac{n_l !}{\left(\frac{n_l+1}{\bar{\rho}}+\frac{1}{\sin^2\theta}\right)\left(\frac{n_l}{\bar{\rho}} + \frac{1}{\sin^2\theta}\right) \cdots \left(\frac{1}{\bar{\rho}}+\frac{1}{\sin^2\theta}\right)}.
	\label{eq_g_func}
	\end{align}
	Hence, the ratio between $g_k^\textrm{channel}\left(n_k\right)$ and $g_j^\textrm{channel}\left(1\right)$ is represented by
	\begin{align}
	&\frac{g_k^\textrm{channel}\left(n_k\right)}{g_j^\textrm{channel}\left(1\right)}=
	\frac{\bar{\rho}^{-n_k} \times q_k \int_0^{\frac{\pi}{2}} \frac{1}{\sin^2\theta} G\left(n_k,\sin^2\theta,\bar{\rho}\right) d\theta}
	{\bar{\rho}^{-1} \times q_j\int_0^{\frac{\pi}{2}} \frac{1}{\sin^2\theta} \left(\left(\frac{2}{\bar{\rho}}+\frac{1}{\sin^2\theta}\right)\left(\frac{1}{\bar{\rho}}+\frac{1}{\sin^2\theta}\right)\right)^{-1} d\theta}.\label{eq_ch_ch_ratio}
	\end{align}
	Since numerator and denominator in \eqref{eq_ch_ch_ratio} have $-n_k$ order of $\bar{\rho}$ and $-1$ order of $\bar{\rho}$ for high SNR regime, respectively, the limit becomes
	\begin{align}
	\lim_{\bar{\rho} \rightarrow \infty} \frac{g_k ^\textrm{channel}\left(n_k\right)}{g_j^\textrm{channel}\left(1\right)} =\lim_{\bar{\rho} \rightarrow \infty} \frac{1}{\bar{\rho}^{n_k-1}}= 0
	\end{align}
	Therefore, $g_k^\textrm{channel}\left(n_k\right) \in o\left(g_j^\textrm{channel}\left(1\right)\right) $ for any $j,k$ such that $n_j = 1$ and $n_k\ge2$.

	Similarly, using \eqref{def_f_gain} and \eqref{eq_del_p_0}, the ratio between $g_k^\textrm{channel}\left(n_k\right)$ and $g_j^\textrm{file}$ is represented by
\begin{align}
\frac{g_k^\textrm{channel}\left(n_k\right)}{g_j^\textrm{file}}=
\frac{\bar{\rho}^{-n_k} \times q_k \int_0^{\frac{\pi}{2}} \frac{1}{\sin^2\theta} G\left(n_k,\sin^2\theta,\bar{\rho}\right) d\theta}
{\bar{\rho}^{-1} \times \frac{q_j}{2} F\left(\bar{\rho}\right)}, \label{eq_ch_f_ratio}
\end{align} where $ F\left(\bar{\rho}\right) =  \left(1+\frac{\beta}{\bar{\rho}} \right)\sqrt{1+\frac{1}{\bar{\rho}}} + \left(1+\frac{1}{\bar{\rho}}\right)\sqrt{1+\frac{\beta}{\bar{\rho}}}.$\\	 
Since numerator and denominator in \eqref{eq_ch_f_ratio} have $-n_k$ order of $\bar{\rho}$ and $-1$ order of $\bar{\rho}$ for high SNR regime, respectively, the limit becomes
\begin{align}
\lim_{\bar{\rho} \rightarrow \infty} \frac{g_k ^\textrm{channel}\left(n_k\right)}{g_j^\textrm{file}} =\lim_{\bar{\rho} \rightarrow \infty} \frac{1}{\bar{\rho}^{n_k-1}}= 0.
\end{align}
Therefore, $g_k^\textrm{channel}\left(n_k\right) \in o\left(g_j^\textrm{file}\right) $ for any $j,k$ such that $n_j = 0$ and $n_k\ge2$.
\section{Proof of Proposition \ref{proposition_4}}\label{pf_prop_4}
\setcounter{equation}{0}
\renewcommand{\theequation}{G.\arabic{equation}}
	To prove the proposition \ref{proposition_4}, we first show the special character of optimal caching placement: as $\bar{\rho} \rightarrow \infty$, the optimal caching placement $\mathbf{n_{\textrm{opt}}}$ satisfies $\left(\left(\mathbf{n_{\textrm{opt}}}\right)_i,\left(\mathbf{n_{\textrm{opt}}}\right)_{N-i+1}\right) \in \left\lbrace \left(1,1\right),\left(2,0\right)\right\rbrace$ for $i \leq \left\lfloor \frac{N}{2} \right\rfloor$, where $\left(\mathbf{n_{\textrm{opt}}}\right)_i$ is the number of helpers that cache the $i$-th popular file.
		By Lemma \ref{lemma_2},  optimal caching placement  consists of $n_i \in \lbrace 0,1,2 \rbrace$. Because $i < N-i+1$ for $i \leq \left\lfloor \frac{N}{2} \right\rfloor$, from Proposition \ref{proposition_1},
		$
		n_i \geq n_{N-i+1}$ for $i \leq \left\lfloor \frac{N}{2} \right\rfloor.$ Then,  for optimal caching placement, feasible combinations of $\left(n_i , n_{N-i+1} \right)$ become $\left\lbrace 			\left(0,0\right),\left(1,0\right), \left(1,1\right),\left(2,0\right),\left(2,1\right) \right\rbrace$.	
		First, we  consider  the case when $\left(n_i, n_{N-i+1}\right) \in  \left\lbrace \left(0,0\right), \left(1,0\right)\right\rbrace$. In this case, by Lemma \ref{lemma_2}, $n_l \leq 2$ for $l \leq i-1$. Also, when $n_{i} \leq 1$, by Proposition \ref{proposition_1}, $n_l \leq 1$ for $l \geq i$, and $n_l = 0$ for $ l \geq N-i+1$ since $n_{N-i+1}=0$. Therefore, the total number of helpers is bounded as
		$
		\sum_{l=1}^F n_l \leq N-1.
		$ This bound implies existence of at least one empty helper, which contradicts the fact that the optimal placement uses up all of helpers' memories. Consequently, the case when $\left(n_i,n_{N-i+1}\right) \in  \left\lbrace \left(0,0\right), \left(1,0\right)\right\rbrace$ cannot lead optimal caching placement.	 
		Second, let us consider the case when $\left(n_i,n_{N-i+1}\right)=\left(2,1\right)$. 
		Then, by Proposition \ref{proposition_1}, $n_l =2$ for $l \leq i$, and $n_l \geq 1$ for $ i+1 \leq l \leq N-i+1$. Using these bound, the total number of helpers is bounded below by
		$
		\sum_{l=1}^F n_l \geq N+1.
		$ This lower bound implies that the caching placement requires at least $N+1$ helpers, which contradicts the fact that we have $N$ caching helpers. Consequently, the case when $\left(n_i, n_{N-i+1}\right) = \left(2,1\right)$ is not feasible. As a result, the optimal caching placement satisfies that $\left(n_i,n_{N-i+1}\right) \in \left\lbrace \left(1,1\right),\left(2,0\right)\right\rbrace$ for $ i \leq \left\lfloor \frac{N}{2} \right\rfloor$.\\
	Using the above property, optimality of the doubly caching placement can be proven as follows. For each $k \in \left\lbrace 1,...,\left\lfloor \frac{N}{2} \right\rfloor \right\rbrace$, we will prove there exists an unique range of Zipf exponent which makes doubly caching placement with $k$ optimal. 
	From the fact that we have just shown, 
	as $\bar{\rho} \rightarrow \infty$,  optimal caching placement has to satisfy
	$
	\left(n_k,n_{N-k+1}\right) = \left(1,1\right) \textrm{ or } \left(2,0\right),$ for $k \leq \left\lfloor \frac{N}{2} \right\rfloor.
	$
	Therefore, optimal caching placement will select a pair between $\left(1,1\right)$ and $\left(2,0\right)$ which gives a higher gain.
	The total gains from each pair are given, respectively, as
	$
	g_k^{\textrm{file}} +  g_{N-k+1}^{\textrm{file}}$ for $\left(n_k,n_{N-k+1}\right) = \left(1,1\right)$, $g_k^{\textrm{file}} + g_k^{\textrm{channel}}\left(1\right)$ for  $\left(n_k,n_{N-k+1}\right) = \left(2,0\right)$.
	Since $g_k^{\textrm{file}}$ is a common term for both of $\left(1,1\right)$ and $\left(2,0\right)$, whether $g_{N-k+1}^{\textrm{file}}$ is larger than $g_k^{\textrm{channel}}\left(1\right)$ or not determines if $\left(n_k,n_{N-k+1}\right) = \left(1,1\right)$ or $\left(n_k,n_{N-k+1}\right) = \left(2,0\right)$ for optimal caching placement.	
	For each $k \in \left\lbrace 1,...,\left\lfloor \frac{N}{2} \right\rfloor \right\rbrace$, if $\gamma \geq \gamma_2\left(k\right)$, 
	\begin{align}
	g_k^{\textrm{channel}}\left(1\right) \geq g_{N-k+1}^{\textrm{file}}.
	\label{ineq_ch_f}
	\end{align}
	Furthermore, using $g_i^{\textrm{channel}}\left(1\right) \geq g_k^{\textrm{channel}}\left(1\right)$ for $i \leq k$ and $g_{N-k+1}^{\textrm{file}} \geq g_l^{\textrm{file}}$ for $l \geq N-k+1$, inequality \eqref{ineq_ch_f} leads to $g_i^{\textrm{channel}}\left(1\right) \geq g_l^{\textrm{file}}$ for $i \leq k$ and $l \geq N-k+1$ which implies that $\left(n_i,n_{N-i+1}\right) = \left(2,0\right)$ is preferred to $\left(n_i,n_{N-i+1}\right) = \left(1,1\right)$ for $i \leq k$. Consequently, for each $k \in \left\lbrace 1,...,\left\lfloor \frac{N}{2} \right\rfloor \right\rbrace$, optimal caching placement satisfies
	\begin{align}
	\left(\mathbf{n_{\textrm{opt}}}\right)_i = 2 \hspace{40pt} \forall i \leq k .
	\label{eq_n_2}
	\end{align}	
	Now, let us consider $\gamma \leq \gamma_3\left(k\right)$ for each $k \in \left\lbrace 1,...,\left\lfloor \frac{N}{2} \right\rfloor-1 \right\rbrace$. We can rewrite $\gamma \leq \gamma_3\left(k\right)$ as
	\begin{align}
	g_{k+1}^{\textrm{channel}}\left(1\right) \leq g_{N-k}^{\textrm{file}}.
	\label{ineq_ch_f_2}
	\end{align}        	
	In contrast to \eqref{ineq_ch_f}, inequality \eqref{ineq_ch_f_2} implies that optimal caching placement satisfies $\left(n_{k+1},n_{N-k}\right) = \left(1,1\right)$ for $k \leq \left\lfloor \frac{N}{2} \right\rfloor-1$. Then, from Proposition \ref{proposition_1},
	\begin{align}
	\left(\mathbf{n_{\textrm{opt}}}\right)_i \leq 1 \hspace{40pt} \forall i \geq k+1 .
	\label{ineq_n_1}
	\end{align}
	From \eqref{eq_n_2} and \eqref{ineq_n_1}, we can conclude that doubly caching placement with $k<\left\lfloor \frac{N}{2} \right\rfloor$ is optimal if $\gamma_2\left(k\right) \leq \gamma \leq \gamma_3\left(k\right)$.\\ 
	However, when $k = \left\lfloor \frac{N}{2} \right\rfloor$, inequality \eqref{ineq_ch_f_2} is equivalent to 
	$\gamma \geq \gamma_3\left(\left\lfloor \frac{N}{2}\right\rfloor\right)$, where $\gamma_3\left(\left\lfloor \frac{N}{2} \right\rfloor \right) < 0$. Obviously, when there are  $N$ helpers, caching $\left\lfloor \frac{N}{2} \right\rfloor +1$-th file in two helpers 
	violates Proposition \ref{proposition_1}. To satisfy Proposition \ref{proposition_1}, \eqref{ineq_n_1} is required when $k = \left\lfloor \frac{N}{2} \right\rfloor$.
	As a result, doubly caching placement with $k = \left\lfloor \frac{N}{2} \right\rfloor$ is optimal if $\gamma \geq\gamma_2\left(\left\lfloor \frac{N}{2} \right\rfloor\right)$. \\	
	 For the converse, we prove that doubly caching placement with $k \left(< \left\lfloor \frac{N}{2} \right\rfloor\right)$ is optimal only if $\gamma_2\left(k\right) \leq \gamma \leq \gamma_3\left(k\right)$ and doubly caching placement with $k = \left\lfloor \frac{N}{2} \right\rfloor$ is optimal only if $\gamma \geq \gamma_2\left(\left\lfloor \frac{N}{2} \right\rfloor\right)$. In order to prove the converse, we prove their contrapositions that if $\gamma < \gamma_2\left(k\right)$ or $\gamma > \gamma_3\left(k\right)$, then doubly caching placement with $k < \left\lfloor \frac{N}{2} \right\rfloor$ is not optimal and if $\gamma < \gamma_2\left(\left\lfloor \frac{N}{2} \right\rfloor\right)$, doubly caching placement with $k = \left\lfloor \frac{N}{2} \right\rfloor$ is not optimal, respectively. \\	
	When $k \in \left\lbrace 1,\cdots,\left\lfloor \frac{N}{2}\right\rfloor \right\rbrace$, $\gamma < \gamma_2\left(k\right)$ is equivalently given by
	$g_k^{\textrm{channel}}\left(1\right) < g_{N-k+1}^{\textrm{file}}$ which means optimal caching placement satisfies $\left(n_k,n_{N-k+1}\right) = \left(1,1\right)$ rather than $\left(n_k,n_{N-k+1}\right) = \left(2,0\right)$. Consequently, optimal caching placement is to cache the $k$-th popular file in a single helper. On the other hand, for $k \in \left\lbrace 1,\cdots,\left\lfloor \frac{N}{2}\right\rfloor -1 \right\rbrace$, $\gamma > \gamma_3\left(k\right)$ can be rewritten as
	$g_{k+1}^{\textrm{channel}}\left(1\right) > g_{N-k}^{\textrm{file}}$ which indicates that optimal caching placement is to cache the  $k+1$-th popular file in two helpers. 
	Hence, if $\gamma < \gamma_2\left(k\right)$ or $\gamma > \gamma_3\left(k\right)$, optimal caching placement is to cache the $k$-th popular file in a single helper or to cache the $k+1$-th popular file in two helpers. Therefore,  doubly caching placement with $k < \left\lfloor \frac{N}{2} \right\rfloor$ is not optimal if $\gamma < \gamma_2\left(k\right)$ or $\gamma > \gamma_3\left(k\right)$ and doubly caching placement with $k =  \left\lfloor \frac{N}{2} \right\rfloor$  is not optimal if $\gamma < \gamma_2\left(\left\lfloor \frac{N}{2} \right\rfloor\right)$.	
\section{Proof of Proposition \ref{proposition_5}}\label{pf_prop_5}
\setcounter{equation}{0}
\renewcommand{\theequation}{H.\arabic{equation}}
	This proposition is proved by contradiction. Suppose doubly caching placement with $k=\lfloor\frac{N}{2}\rfloor$ is not optimal. Then, there exists another optimal caching placement. Let us denote optimal caching placement as $\mathbf{n_{\textrm{opt}}}$ and doubly caching placement with $k=\left\lfloor \frac{N}{2} \right\rfloor$ as $\mathbf{n_{\left\lfloor \frac{N}{2} \right\rfloor}}$. Since Proposition \ref{proposition_4} characterizes optimal caching placement in whole range of $\gamma$ in high SNR regime, $\mathbf{n_{\textrm{opt}}}$ has to be doubly caching placement with $k <\left\lfloor\frac{N}{2}\right\rfloor$ for some $k$.
	If we subtract the average BER of $\mathbf{n_{\left\lfloor \frac{N}{2} \right\rfloor}}$ from the that of $\mathbf{n_\textrm{opt}}$, the difference must be less than zero due to optimality, i.e., 
	$
	p_e\left(\mathbf{n}_{\textrm{opt}} \right) - p_e \left( \mathbf{n}_{\lfloor\frac{N}{2}\rfloor}\right) < 0.
	$	
	Now, we show the difference becomes larger than zero. The difference of the average BER is expressed as
	\begin{align}
	p_e\left(\mathbf{n}_{\textrm{opt}} \right) - p_e \left( \mathbf{n}_{\lfloor\frac{N}{2}\rfloor}\right) &= \sum_{i=1}^F q_i\left(p_{e}\left(\left(\mathbf{n}_{\textrm{opt}}\right)_i\right) - p_{e}\left(\left(\mathbf{n}_{\lfloor\frac{N}{2}\rfloor}\right)_i\right)\right) \\
	&\overset{(a)}{=} - \sum_{i \in \mathcal{F}_1} g^{\textrm{file}}_i + \sum_{i \in \mathcal{F}_2} g^{\textrm{channel}}_i\left(1\right)  \\
	&\overset{(b)}{>} \sum_{i \in \mathcal{F}_2} \left(g^{\textrm{channel}}_i\left(1\right) - g^{\textrm{file}}_i \right)\\
	& \overset{(c)}{=} \sum_{i \in \mathcal{F}_2}  \frac{q_i}{2}\left(\sqrt{\frac{\bar{\rho}}{\beta+\bar{\rho}}} - \sqrt{\frac{\bar{\rho}}{2+\bar{\rho}}} \right), 
	\label{ineq_erro_dif}
	\end{align}where $\mathcal{F}_1$ is a set of files cached in $\mathbf{n_{\textrm{opt}}}$ but not cached in $\mathbf{n}_{\left\lfloor \frac{N}{2} \right\rfloor}$, and  $\mathcal{F}_2$ is a set of files  cached doubly in $\mathbf{n}_{\left\lfloor \frac{N}{2} \right\rfloor}$ but not doubly cached in $\mathbf{n_{\textrm{opt}}}$, respectively. 
	Except for the files in  $\mathcal{F}_1$ and $\mathcal{F}_2$, other elements are the same in $\mathbf{n_{\textrm{opt}}}$ and $\mathbf{n_{\left\lfloor \frac{N}{2} \right\rfloor}}$. Therefore, we can express the difference as a function of channel diversity gain and file diversity gain  as $(a)$; to be cached only in $\mathbf{n_{\textrm{opt}}}$, the popularity of the file has to be less than $\left\lfloor \frac{N}{2} \right\rfloor$. Also, the files which has lower popularity than $\left\lfloor \frac{N}{2} \right\rfloor$ cannot be cached doubly due to Proposition \ref{proposition_1} for optimal caching placement. Thus, all the files in $\mathcal{F}_1$ are less popular than all the files in $\mathcal{F}_2$. If we increase the popularity of the files in $\mathcal{F}_1$ up to the popularity of $\mathcal{F}_2$, we can make a lower bound $(b)$; Applying \eqref{eq_del_p_0} and \eqref{eq_del_p_1} to the definitions of both diversity gain \eqref{def_f_gain} and \eqref{def_ch_gain}, we can get $(c)$. 
	\eqref{ineq_erro_dif} is greater than zero due to $\beta <2$, which contradicts the supposition that doubly caching placement with $k=\lfloor\frac{N}{2}\rfloor$ is not optimal.

\ifCLASSOPTIONcaptionsoff
  \newpage
\fi



%

\end{document}